# A Data-driven Long-term Dynamic Rating Estimating Method for Power Transformers

Ming Dong, *Senior Member, IEEE*

*Abstract*—This paper presents a data-driven method for estimating annual continuous dynamic rating of power transformers to serve the long-term planning purpose. Historically, research works on dynamic rating have been focused on real-time/near-future system operations. There has been a lack of research for long-term planning oriented applications. Currently, most utility companies still rely on static rating numbers when planning power transformers for the next few years. In response, this paper proposes a novel and comprehensive method to analyze the past 5-year temperature, loading and load composition data of existing power transformers in a planning region. Based on such data and the forecasted area load composition, a future power transformer's load shape profile can be constructed by using Gaussian Mixture Model. Then according to IEEE std. C57.91-2011, a power transformer thermal aging model can be established to incorporate future loading and temperature profiles. As a result, annual continuous dynamic rating profiles under different temperature scenarios can be determined. The profiles can reflect the long-term thermal overloading risk in a much more realistic and granular way, which can significantly improve the accuracy of power transformer planning. A real utility application example in Canada has been presented to validate and demonstrate the practicality and usefulness of this method.

## I. INTRODUCTION

ACCURATE long-term planning is the key to ensure balanced cost and reliability of power system in the next 5-10 years. As a critical and costly component, power transformer planning is an important part of long-term system planning process, in which the forecasted area load to be supplied by the transformer is compared with transformer's rating to determine the proper transformer sizing.

However, most utility companies currently use static power transformer rating assumption, in many cases the nameplate ratings for long-term system planning [1-4].These assumptions can be overly conservative or inaccurate as they do not reflect the dynamic temperature conditions in the planning region throughout a year. This is especially true for relatively cold areas such as Canada where the ambient temperatures are relatively low. According to IEEE std. C57.91-2011, the insulation deterioration of power transformers is a function of dynamic loading and ambient temperature. Proper combinations of dynamic loading and ambient temperature could safely allow transformer loading to exceed the nameplate rating without causing any damage. Therefore, to improve the cost-effectiveness of planning decisions, a scientific and realistic way to establish annual continuous dynamic rating for power transformers is required.

Previously, research works on dynamic rating mainly focused on real-time or near-future operations of system equipment [5-9]. Based on the monitoring of electrical and environmental conditions, real-time or near-future equipment ratings can be estimated or predicted and flexible loading operations or asset management decisions can be optimized accordingly to capitalize on such varying ratings. The research on establishing typical annual dynamic ratings to serve the long-term planning purpose has not been found. For such applications, there are two unique challenges:

1) No monitoring data is available for long-term future. Since the purpose of planning is to study the future load growth of an area, both long-term loading and temperature profiles are currently unknown and have to be estimated. Also, due to the high uncertainties over a long-term planning horizon, different scenarios may need to be studied.

2) Unlike operational dynamic rating which usually focuses on a short period of time such as a few hours or a few days, dynamic rating for long-term planning should be established on an annual basis to cover different seasons.

To tackle the above challenges, this paper proposes a novel and comprehensive data analytics method as shown in Fig.1. Each step in the flowchart is explained as follows:

- Step 1: The past 5-year hourly temperature data in the planning region is analyzed to establish three long-term annual temperature profiles under three scenarios;
- Step 2: For each future day in the 365-day profile, 5 historical days that have closest temperature and calendar characteristics are found;
- Step 3: Within these 5 days, the relationships between the existing transformers' load compositions and the future transformer's forecasted load composition are analyzed by using Gaussian Mixture Model and Silhouette analysis in a probabilistic way;
- Step 4: By incorporating 24-hr loading profiles of existing transformers and the probabilistic relationships established in Step 3, the future transformer's normalized load shape profile can be constructed;

M. Dong is with Department of System Planning and Asset Management, ENMAX Power Corporation, Calgary, AB, Canada, T2G 4S7 (e-mail: mingdong@ieee.org)

- Step 5: In the last step, the load shape profile along with the established 24-hr ambient temperature profile are fed into the transformer thermal aging model established according to IEEE std. C57.91-2011. The normalized load shape profile is proportionally scaled up until accelerated transformer aging starts to appear. At this point, the power transformer's dynamic rating for this particular profile day is determined since accelerated aging should be avoided for long-term power asset investment.

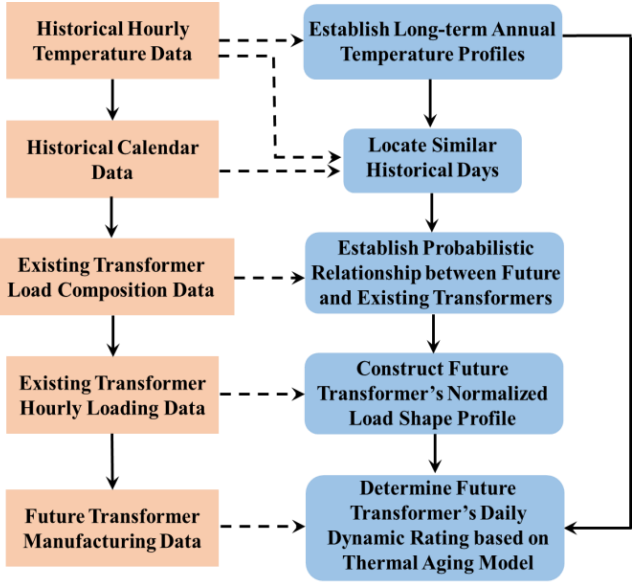

Fig. 1. Flowchart of the proposed data analytics method

Repeat steps 2-5 until the ratings for all 365 days under the three temperature scenarios established in Step 1 are determined. The established annual dynamic rating profiles can reflect the long-term thermal overloading risk in a much more realistic and granular way, which can significantly improve the cost-effectiveness of power transformer planning.

In the following sections, this paper explains each step above in detail. In the end, a real application example in a utility company in West Canada is given to present the established annual dynamic rating profiles. A sensitivity study is also given to demonstrate how the results vary with the forecasted composition of area loads to be supplied by the future transformer. In summary, this paper presents a unique method of establishing annual continuous dynamic rating, for long-term power transformer planning purpose.

## II. LONG-TERM ANNUAL TEMPERATURE PROFILING AND SIMILAR HISTORICAL DAYS

This section explains the details of step 1 and 2 in the flowchart of Fig.1. First, the process of establishing long-term temperature profiles under different scenarios is discussed; second, the method of finding 5 closest historical days based on temperature and calendar features is given.

### A. Establishing Long-term Annual Temperature Profiles

It is a basic fact that long-term hourly temperature profiles cannot be accurately forecasted [10]. However, given the past 5-year temperature data in a planning region such as a city or a town, the statistically representative long-term temperature profiles can be established. Three temperature scenarios high, medium and low are considered for planning purpose. In the high temperature scenario, for each day in the 365 days, the average daily temperatures in the past 5 years are compared and the day under the year with the highest average daily temperature is selected. For example, to create a profile for January 1st, January 1st in the past 5 years are compared by average daily temperature and it is found that 2016 January 1st has the highest daily temperature. Then the 24-hr temperature profile of 2016 January 1st is selected under the highest temperature scenario. This process continues until all 365 days' profiles are selected from history and concatenated. Medium and low temperature scenarios use the same process except that when comparing among 5 years, instead of selecting the highest daily temperature day, the days with median and lowest daily temperatures are selected.

In addition to the above selection and concatenation process, a safety margin or global warming adjustment such as 1°C can be artificially added to all profiles. In this case, every hour under the three scenarios will be increased by 1°C.

The above method is unique in the sense that on the one hand, it reflects the future temperatures at three levels; on the other hand, it keeps the authentic temperature pattern within each day: each profile day has a corresponding historical day in the past 5 years and hence has a high creditability.

### B. Locating Similar Historical Days through Comparison

The next step is to find 5 similar historical days for each profile day established in subsection A. The purpose of this step is to find proper days based on which advanced data analytics can be further applied to construct the transformer's load shape profile, as to be discussed in Section III. To find similar days, two groups of features are considered: temperature and calendar features.

1) Temperature features: as [11-15] suggest, temperature can significantly affect the loading behaviour. For example, air conditioning is more frequently used in hot days and the consumed power demand has a positive correlation with the ambient temperature. Ambient temperature may also affect customer behaviours since customers tend to stay indoor when it is very cold or hot outside and this behaviour often lead to increased power usage. To characterize daily temperatures, maximum, average and minimum temperatures in a day are chosen as features. For a 24-hr temperature profile, they are:

$$\begin{cases} T_a = \dfrac{\sum_{i=1}^{24} T_i}{24} \\ T_h = max(T_1, T_2, ..., T_{24}) \\ T_l = min(T_1, T_2, ..., T_{24}) \end{cases} \quad (1)$$

where $T_1$ to $T_{24}$ are the hourly temperatures in a day.

2) Calendar features: as [14-15] suggest, workdays and holidays including weekends could have significantly different loading patterns. For example, in general, residential customers consume more power on weekends and industrial customers consume more power on weekdays. Therefore, it is important to separate workdays and holidays into two groups and search for similar days within the two groups respectively.

Another introduced calendar feature is to reflect the position of a day in the annual cycle, i.e. day of the year. This feature could also imply different loading patterns. For example, although a major industrial load on two workdays in the Fall and Spring have similar temperatures, it has significantly different loading patterns at two very different times of a year. By using day of the year feature, the numerical difference on the yearly calendar can be reflected. According to [16], the day of the year feature can be mathematically defined as:

$$Y = sin\left(2\pi * \frac{D}{365}\right) \quad (2)$$

where $D$ is the day in 365 days. For example, $D$ for January 1[st] is 1 and $D$ for December 31[st] is 365. Sine function is used is to reflect the cyclic characteristic and avoid one-way increase of the numerical value $D$.

To locate 5 historical days with similar temperature and calendar features, at the beginning workdays and holidays are separated into two different groups due to significant distinctions between them. Then within each group, similar to many clustering analysis methods that rely on Euclidean distance to measure the differences between data points [17], this paper proposes to use the following Euclidean distance formula to measure the distances between a historical day $D$ and the targeted profile day $D'$.

$$d = \sqrt{(T_a - T_a')^2 + (T_h - T_h')^2 + (T_l - T_l')^2 + (Y - Y')^2} \quad (3)$$

where $T_a$, $T_h$, $T_l$ and $Y$ are the temperature and day of the year features of the historical day $D$; $T_a'$, $T_h'$ and $Y'$ are the temperature and day of the year features of the profile day $D'$. It should be noted that before applying (3), the features are all normalized to [0,1] by using (4) to eliminate magnitude and unit differences. The following equation can be used for normalization [17]:

$$f_{norm} = \frac{f_{raw} - min(f)}{max(f) - min(f)} \quad (4)$$

where $max(f)$ is the maximum value observed in the feature $f$ and $min(f)$ is the minimum value observed in feature $f$; $f_{raw}$ is the raw value of the temperature or calendar feature.

In the end, 5 historical similar days with minimum distances measured by (3) are selected out of the past 5 years and form the data windows for further analytics to be applied as discussed in the following sections.

## III. FUTURE TRANSFORMER LOAD SHAPE PROFILING

This section explains the details of Step 3 and 4 in the flowchart of Fig.1. The ultimate goal is to create the normalized 24-hr load shape profile for the future transformer for a specific profile day in 365 days. An important concept called "Transformer Load Composition" is introduced and quantified. This is because the transformer total load is composed of residential, commercial and industrial loads supplied by the transformer. Different types of loads have different load shapes throughout a day and can respond to ambient temperatures in different ways.

In this section, an important probabilistic clustering method Gaussian Mixture Modeling and an efficient clustering quality evaluation method Silhouette analysis are explained. They are used together to quantify the probabilistic relationship between the future transformer and existing transformers based on transformer load composition. Based on the probabilistic clustering result, the normalized load shape profile for the future transformer can be constructed based on weighted average.

### A. Transformer Load Composition

In general, most power transformers supply more than one type of loads. Approximately, the loads can be categorized into three types: residential, commercial and industrial loads. Transformer load composition can be described by the percentages of every load type. Residential load percentage $R$, commercial load percentage $C$ and industrial load percentage $I$ should comply with:

$$R + C + I = 1 \quad (5)$$

When a customer load is connected or planned to be connected to a utility grid, it is a common practice for utility companies to assign the load to the above three categories with different electricity rates. Therefore, $R$, $C$ and $I$ can be easily determined. If needed, sub-categories of commercial and industrial loads can be determined on an individual load basis. However, this would require heavy manual classification work by human experts. In such a case, (5) becomes:

$$R + \sum_{i=1}^{m} C_i + \sum_{j=1}^{n} I_j = 1 \quad (6)$$

where there are $m$ pre-determined commercial load subcategories and $n$ pre-determined industrial load subcategories.

For a historical day, $R$ can be calculated as:

$$R = \sum_{i=1}^{m} \frac{L_{i,t}^R}{P_t} \times 100\% \quad (7)$$

where $P_t$ is the transformer peak loading in the day; $m$ is the total number of residential loads supplied by the transformer; $L_{i,t}^R$ is the loading of each residential load $i$ at the transformer peaking time of the day. Similarly to residential load, transformer commercial load percentage is calculated as:

$$C = \sum_{i=1}^{n} \frac{L_{i,t}^C}{P_t} \times 100\% \quad (8)$$

where $L_{i,t}^C$ is the loading of each commercial load $i$ at the transformer peaking time of the day; $n$ is the total number of commercial loads supplied by the transformer.

It should be noted for historical days, the loading values of existing customers in a day can be obtained from interval metering data and $R$ and $C$ can be calculated using (7) and (8); for a new area in a future day, $R$ and $C$ are estimated based on the expected numbers of residential, commercial and industrial customers along with their typical coincidental unit loading. When (5) is used to characterize transformer loading, only two percentage numbers out of the three are required to characterize the load composition. This means the clustering dimensionality can be reduced to 2. For example, if $R$ and $C$ are selected, a power transformer can be characterized simply as a vector $(R, C)$; however when (6) is used, the transformer will need to be characterized with multiple dimensions and the clustering performance may be affected.

## B. Gaussian Mixture Modeling

Unlike deterministic clustering methods such as K-Means and Mean-shift which requires each data point to belong to a single cluster, Gaussian Mixture Model (GMM) is a powerful probabilistic clustering method [18-20]. When using GMM, a data point can belong to all clusters with certain membership probabilities. In statistics, a Gaussian mixture model is a mixture distribution that assumes all the data points are generated from a mixture of a finite number of Gaussian distributions with certain parameters to be determined. For clustering analysis, a Gaussian mixture function is comprised of several Gaussian components i.e. clusters, each identified by $k \in \{1,...,K\}$, where $K$ is the expected number of clusters in the dataset $U$. Each cluster $k$ in the mixture has the following three parameters:

- Mean $\mu_k$ which defines the centroid of cluster $k$;
- Mixture weight $w_k$ which describes how cluster $k$ gets mixed into the global mixture function;
- Covariance matrix $\Sigma$ of cluster $k$. In a $n$-dimensional case, cluster $k$ can be written as a column vector:

$$X = (X_1, X_2, ..., X_n)^T \qquad (9)$$

In the covariance matrix $\Sigma$ shown below:

$$\Sigma = \begin{bmatrix} Cov_{1,1} & \cdots & Cov_{1,n} \\ \vdots & \ddots & \vdots \\ Cov_{n,1} & \cdots & Cov_{n,n} \end{bmatrix} \qquad (10)$$

each matrix element $Cov_{i,j}$ is defined as:

$$Cov_{i,j} = E[(X_i - E[X_i])(X_j - E[X_j])]$$
$$= E[X_iX_j] - E[X_i]E[X_j] \qquad (11)$$

where $E$ is the expected value of its data array argument. In a one-dimensional case, $\Sigma$ has only one element and it is equivalent to the variance of the data points in cluster $k$.

The standard multivariate Gaussian probability density function is mathematically given as below:

$$N(x|\mu, \Sigma) = \frac{1}{(2\pi)^{\frac{n}{2}} |\Sigma|^{\frac{1}{2}}} exp\left[-\frac{1}{2}(x - \mu)^T \Sigma^{-1}(x - \mu)\right] \qquad (12)$$

Gaussian mixture model that consists of $K$ Gaussian components is defined as:

$$p(x) = \sum_{k=1}^{K} w_k \cdot N_k(x|\mu_k, \Sigma_k) \qquad (13)$$

where $w_k$ is the weight of $k_{th}$ Gaussian component and it complies with:

$$\sum_{k=1}^{K} w_k = 1, 0 \leq w_k \leq 1 \qquad (14)$$

For illustration purpose, a one-dimensional Gaussian mixture probability density function that consists of 3 Gaussian distributions $x_1 \sim N(5,4), x_2 \sim N(10,4)$ and $x_3 \sim N(15,4)$ with equal mixing weight 1/3 is plotted in Fig.2.

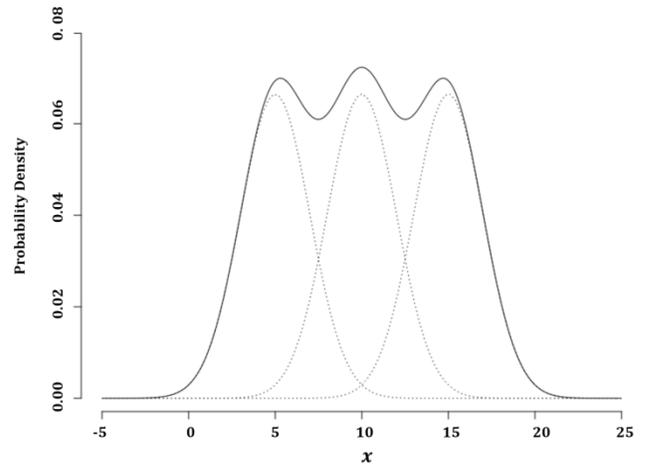

Fig. 2. An example of one-dimensional Gaussian mixture model probability density function

The GMM clustering can be determined by using EM (Expectation-Maximization) algorithm. The EM algorithm consists of the E-step and the M-step: in the beginning, $K$ Gaussian distributions are randomly parameterized and then in the E-step, for each data point, the probability of it belonging to the $K$ Gaussian distributions are calculated by using Bayers' theorem. In the subsequent M-step, the parameters of the Gaussian distributions get updated with the probabilistically associated data points. The E-step and M-step repeat iteratively until convergence is reached. Details of EM algorithm can be found in [21]. After the Gaussian distribution parameters are determined, for a data point $x$ in the dataset $U$, it can simultaneously belong to all $K$ clusters (distributions) with the membership probability $p_{xk}$ for each cluster $k$:

$$p_{xk} = \frac{w_k \cdot N_k(x|\mu_k, \Sigma_k)}{\sum_{k=1}^{K} w_k \cdot N_k(x|\mu_k, \Sigma_k)}, x \in U \qquad (15)$$

$p_{xk}$ is the key parameter used to estimate future transformer's load shape profile and will be further used in subsection D.

## C. Clustering Quality Evaluation using Silhouette Analysis

Although GMM provides a mathematically sound way for probabilistic clustering analysis, the expected number of clusters $K$ is unknown. One way to determine $K$ is evaluating the clustering quality under different $K$ values and selecting $K$ which yields the best clustering quality. In order to evaluate clustering quality, Silhouette analysis is adopted [22]. In this analysis, an index called Silhouette coefficient $Q_r$ is used to evaluate clustering quality. For a given data point ($r \in$ cluster $S_r$), its $Q_r$ can be mathematically calculated using the equations below:

$$\begin{cases} Q_r = \dfrac{b_r - a_r}{max(a_r, b_r)} \\ a_r = \dfrac{1}{|S_r| - 1} \sum_{s \in S_r, r \neq s} d(r, s) \\ b_r = min[\dfrac{1}{|S_v|} \sum_{v \in S_v} d(r, v)] \end{cases} \qquad (16)$$

where $|S_r|$ is the number of members in cluster $S_r$; $S_v$ is any other cluster in the dataset; data point $v$ is data point in $S_v$; $d$ is the Euclidean distance between two data points.

To evaluate the clustering quality, (16) calculates both the compactness and separation of produced clusters by GMM: $a_r$ reflects the intra-cluster compactness. It is the average distance of data point $r$ to all other points in the same cluster $S_r$; $b_r$ reflects the separation between other clusters and point $r$. It is the smallest average distance of $r$ to all points in every other cluster that does not contain $r$ in the dataset; $Q_r$ is the final index that combines $a_r$ and $b_r$. A good intra-cluster compactness and inter-cluster separation together will lead to a large $Q_r$ value.

(16) is the calculation for a single data point $r$. To evaluate the clustering quality of the entire dataset, average Silhouette coefficient is used and is given as below:

$$Q_{avg} = \frac{1}{m} \sum_{i=1}^{m} Q_i \qquad (17)$$

where $m$ is total number of data points in this dataset $U$.

$Q_{avg}$ for an initial range of $K$ values is tested and then the $K$ value resulting in the highest $Q_{avg}$ is selected as the optimal $K$ and used in GMM.

### D. Constructing Normalized Load Shape Profile for the Future Transformer

By using GMM, existing transformers within the 5 days identified in Section II along with the future transformer are clustered together based on their load composition features. An example of clustering result based on residential load percentage $R$ and commercial load percentage $C$ features for 80 transformers in 5 days with 6 clusters is shown in Fig.3. $R$ and $C$ have been normalized by using (4).

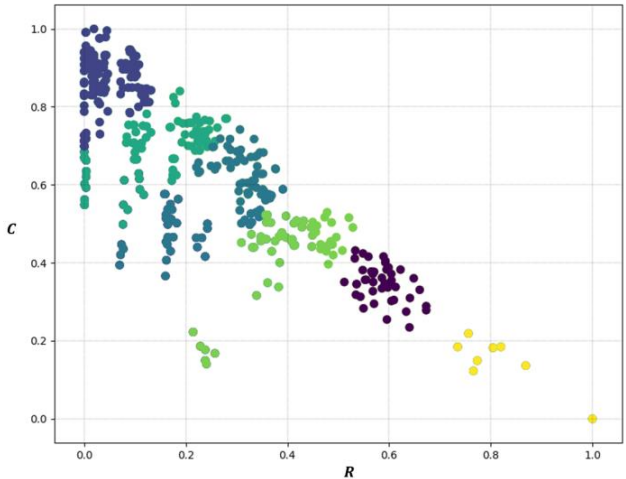

Fig. 3. An example of transformer GMM clustering result

As previously discussed, (15) can be used to calculate the membership probability $p_{xk}$ of the future transformer to each cluster. The future transformer's normalized loading at $t_{th}$ hour $L_{pu}(t)$ can be calculated as:

$$L_{pu}(t) = \sum_{k=1}^{K} p_{xk} \cdot \frac{L_k(t)}{P_k} \qquad (18)$$

where $L_k(t)$ is the loading of cluster centroid $k$ at $t_{th}$ hour; $P_k$ is the peak loading of cluster centroid $k$ in that day.

(18) is based on the principle that if the future transformer's load composition on the profile day is similar to a group of

existing transformers' load compositions on similar historical days, its load shape (reflected as normalized profile) should also be similar to the load shape of such existing transformers. An example of a constructed load shape profile versus normalized loading profiles of 6 cluster centroids is plotted in Fig.4.

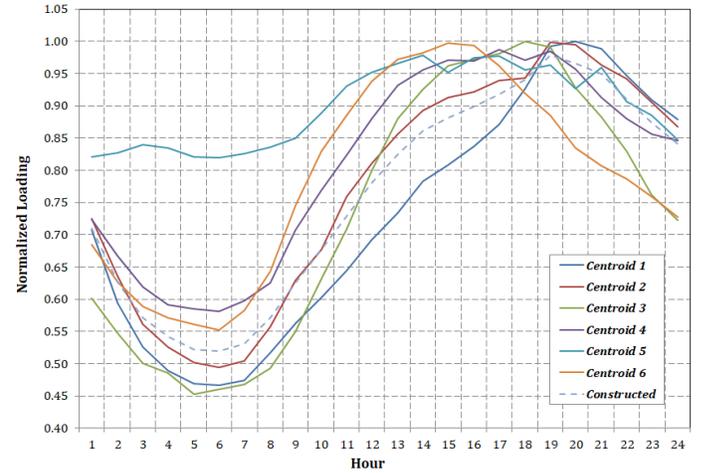

Fig. 4. An example of constructing transformer load shape profile

## IV. POWER TRANSFORMER THERMAL AGING MODEL

This section explains the details of step 5 in the flowchart of Fig.1. IEEE std. C57.91-2011 explains the quantitative relationship between transformer thermal aging and influencing factors such as transformer loading and ambient temperature [22-23]. This section first explains the method to calculate equivalent aging factor and then explains the method to derive transformer dynamic load rating.

### A. Calculate Equivalent Aging Factor

According to IEEE std. C57.91-2011, Fig.5 summarizes the steps to calculate transformer equivalent aging factor: first, transformer top-oil temperature rise over ambient temperature is calculated; second, transformer hottest-spot temperature rise over top-oil temperature is calculated; third, the end of hour hottest-spot temperature is calculated; then the end of hour hottest-spot temperature is converted to transformer hourly aging acceleration factor; in the end, the transformer 24-hr equivalent aging factor is calculated.

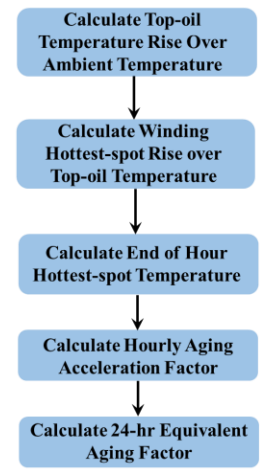

Fig. 5. Flowchart of calculating transformer equivalent aging factor

In the first step, the transformer top-oil temperature rise over ambient temperature is calculated using equations below:

$$\begin{cases} \Delta\Theta_{TO} = \left(\Delta\Theta_{TO,U} - \Delta\Theta_{TO,i}\right)\left(1 - e^{-\frac{24}{\tau_{TO}}}\right) + \Delta\Theta_{TO,i} \\ \Delta\Theta_{TO,U} = \Delta\Theta_{TO,R}\left[\frac{(K_u^2 R + 1)}{(R + 1)}\right]^n \end{cases} \quad (19)$$

where $\Delta\Theta_{TO}$ is the end of hour top-oil rise over ambient temperature in ℃; $\Delta\Theta_{TO,i}$ is the initial top-oil rise over ambient temperature in ℃; $\Delta\Theta_{TO,U}$ is the ultimate top-oil rise over ambient temperature in ℃; $K_u$ is the ratio of current-hour loading to rated loading; $\tau_{TO}$ is the transformer time constant for temperature differential between the ultimate top-oil rise and initial top-oil rise and can be provided by the transformer manufacturer; $\Delta\Theta_{TO,R}$ is a constant representing the top-oil rise over ambient temperature at rated loading on the tap position to be studied and can be provided by the transformer manufacturer; $R$ is a constant representing the ratio of load loss at rated loading to no-load loss and can be provided by the transformer manufacturer; $n$ is an empirical exponent. It is 0.8 for power transformers with natural convection flow of oil and natural convection flow of air over radiators (ONAN type). It is 0.9 for power transformers with natural convection flow of oil and forced convection flow of air over radiators by fans (ONAF type) [23].

It should be noted that when applying (19), the initial top-oil rise over ambient temperature $\Delta\Theta_{TO,i}$ for each hour is unknown. A loop-based iterative calculation process is often used to solve this problem: $\Delta\Theta_{TO,i}$ in the first hour of the day is initialized to a low temperature number such as 0℃. Then $\Delta\Theta_{TO}$ in the first hour is calculated and also used as the input $\Delta\Theta_{TO,i}$ for the second hour. This process continues until values in all 24 hours get calculated. Then $\Delta\Theta_{TO}$ in the last hour is used as input $\Delta\Theta_{TO,i}$ for the first hour. The loop calculation continues until no hourly values get updated and this typically happens after a few iterations.

In the second step, the winding hottest-spot rise over top-oil temperature is calculated by using:

$$\begin{cases} \Delta\Theta_H = \left(\Delta\Theta_{H,U} - \Delta\Theta_{H,i}\right)\left(1 - e^{-\frac{24}{\tau_W}}\right) + \Delta\Theta_{H,i} \\ \Delta\Theta_{H,i} = \Delta\Theta_{H,R}K_i^{2m} \\ \Delta\Theta_{H,U} = \Delta\Theta_{H,R}K_u^{2m} \end{cases} \quad (20)$$

where $\Delta\Theta_H$ is the end of hour winding hottest-spot rise over top-oil temperature in ℃; $\Delta\Theta_{H,i}$ is the initial winding hottest-spot rise over top-oil temperature in ℃; $\Delta\Theta_{H,U}$ is the ultimate winding hottest-spot rise over top-oil temperature in ℃; $K_i$ is the ratio of last-hour loading to rated loading; $\tau_W$ is the winding time constant at hot spot location and can be provided by the transformer manufacturer; $\Delta\Theta_{H,R}$ is a constant representing transformer hotspot differential and can be provided by the transformer manufacturer; $m$ is an empirical factor. It is 0.8 for most power transformers and 1.0 for the ones that direct oil from the radiators or heat exchangers into the windings and force air over the radiators or heat exchanger by fans (ODAF type) [23].

In the third step, the end of hour hottest-spot temperature is calculated by using:

$$\Theta_H = \Theta_A + \Delta\Theta_{TO} + \Delta\Theta_H \quad (21)$$

where $\Theta_A$ is the hourly ambient temperature in ℃.

In the fourth step, according to Arrhenius reaction rate theory, the hourly aging acceleration factor $F_{AA}$ is calculated by using:

$$F_{AA} = e^{\left[\frac{15000}{383} - \frac{15000}{\Theta_H + 273}\right]} \quad (22)$$

In the fifth step, the transformer 24-hr equivalent aging factor $F_{EQA}$ is calculated by using:

$$F_{EQA} = \frac{\sum_{t=1}^{24} F_{AA,t}}{24} \quad (23)$$

where $t$ is the hour in a day.

### B. Determine Transformer Daily Dynamic Rating

From the long-term planning perspective, it is desired that the 24-hr equivalent aging factor $F_{EQA}$ is 1.0. This is because when $F_{EQA}$ is less than 1.0, the power transformer is underutilized against its normal insulation life (underloading situation); when $F_{EQA}$ is greater than 1.0, the power transformer is over-utilized against its normal insulation life and the overall life will be shortened (overloading situation). Therefore, keeping $F_{EQA}$ as one is used as the criterion to determine the daily transformer load rating.

In Section III, the transformer load shape profile has been constructed based on load composition. Since it is normalized, it only captures the load shape and does not reflect the loading magnitude. In this step, the normalized profile is proportionally scaled up with a small step change and at each step, the corresponding $F_{EQA}$ gets calculated until $F_{EQA} = 1$ is reached. An example of a 50MVA power transformer's 24-hr thermal aging simulation during a day is shown in Fig.6. In this example, $F_{EQA} = 1$ and as can be seen, a significant portion of the transformer load $K_u$ is greater than rated loading 1.0 p.u. The maximum transformer load during the day is actually 1.55 p.u. This means the transformer dynamic rating for the day is 77.5MVA, for the particular load composition and temperature profile in this example.

### Power Transformer 24-hr Thermal Aging Simulation

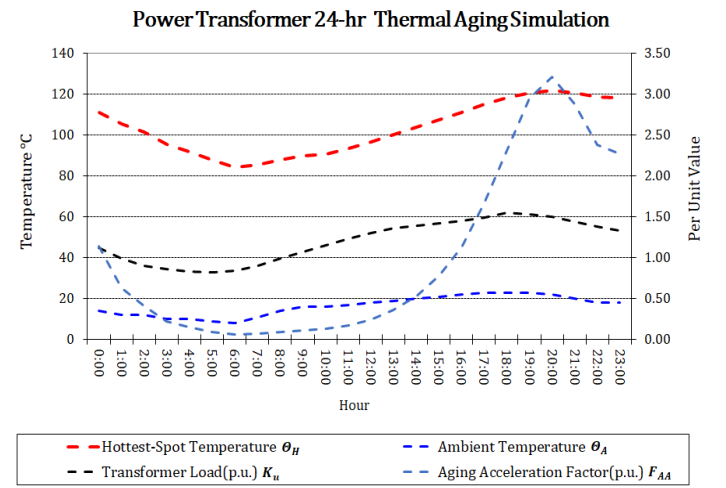

Fig. 6. An example of power transformer 24-hr thermal aging simulation

## V. VERIFICATION AND APPLICATION

The proposed method has been applied to a major utility company in West Canada for one of its planning regions in the Alberta Province. The results of verification and application are presented and discussed in detail in this section.

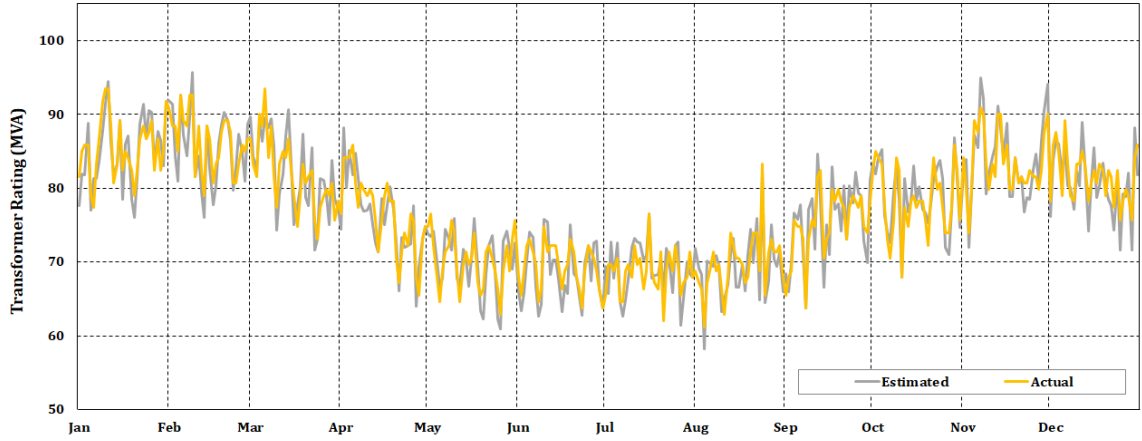

Fig. 7. T1's 2018 Actual Dynamic Rating vs. Estimated Dynamic Rating (with actual temperature profile)

### A. Verification with Actual Temperature Profile

To better observe and analyze the details of the proposed power transformer long-term dynamic rating estimating method, this paper adopts a two-step verification for the proposed comprehensive method. In the first step, the focus of verification is the load shape profiling method discussed in Section III: 5 representative power transformers (T1 to T5) out of the total 80 power transformers in the planning region were chosen and used for verification. Data from 2013 to 2018 were used. The 5-year data from year 2013 to 2017 is used to estimate the annual dynamic rating in 2018. The 2018 daily temperature data is treated as known data. The 5 transformers' normalized load shape profiles are constructed by using the proposed method. They are then taken into the IEEE power transformer thermal aging model and scaled up gradually to determine their estimated daily dynamic rating values. In comparison, the actual 2018 daily loading profiles of the 5 transformers are scaled in the same way to determine their actual daily dynamic rating values.

The 2018 actual and estimated dynamic rating values of transformer T1 are plotted in Figure 7. As can be seen, although there are gaps between the two rating curves on individual days, the two curves follow a quite consistent pattern and stay around the same level. Three error metrics are used to quantify the estimation accuracy:

1) For daily error, Mean Absolute Percentage Error (*ME*) is used and it is mathematically given as below [26]:

$$ME = \frac{1}{N} \sum_{i=1}^{N} \left| \frac{R_i^A - R_i^E}{R_i^A} \right| \times 100\% \qquad (24)$$

where $N$ is the total number of days in a forecasting period; $R_i^A$ is the actual dynamic rating of the $i_{th}$ day and $R_i^E$ is the estimated dynamic rating of the $i_{th}$ day.

2) Since the task discussed in this paper is for the purpose of long-term economic transformer sizing, what is more important than the daily accuracy is the statistical accuracy during a forecasting period such as summer forecasting season (May to Sep.) and winter forecasting season (Oct. to Apr.) in one year. Therefore, in addition to using *ME*, two statistical metrics are proposed in order to describe the error of the estimation for a specific forecasting period: Average Rating

Percentage Error (*AE*) and Valley Rating Percentage Error (*PE*). Mathematically, *AE* and *PE* are defined as below:

$$AE = \frac{|\overline{R_t^A} - \overline{R_t^E}|}{\overline{R_t^A}} \times 100\% \qquad (25)$$

$$VE = \frac{|min(R_i^A) - min(R_i^E)|}{min(R_i^A)} \times 100\% \qquad (26)$$

where $\overline{R_t^A}$ and $\overline{R_t^E}$ are the average values of actual and estimated dynamic ratings in the forecasting period; $min(R_i^A)$ and $min(R_i^E)$ are the minimum values of actual and estimated daily dynamic ratings in the forecasting period. *AE* describes the error of the average daily rating and *VE* describes the error of the minimum daily rating. Compared to the peak point, the valley point *VE* is more useful because it reflects the minimum rating required for the transformer.

The daily dynamic ratings of the pre-selected five transformers in 2018 are estimated and the errors are further calculated by using equation (24)-(26). The results are summarized into Table I. Furthermore, to facilitate the result analysis, the average load composition forecasted for 2018 and the average membership probability for each transformer are also included in the table. In Section III, it has been discussed that for each profile day, a few historical similar days can be found and then GMM based probabilistic clustering is applied to these similar days. For each profile day, the outcome of this step is that the target transformer can be associated to $K$ clusters of existing transformers and the association can be quantified by a set of membership probability numbers $p_{xk}$ ($k$=1 to $K$). The average membership probability listed in Table I is the annual average of every day's maximum membership probability among the $K$ probability numbers.

As shown in Table I, 5 representative transformers are selected for discussion purpose. These transformers cover the typical range of load compositions in the planning region: T1 is commercial heavy; T2 is residential and commercial heavy; T3 is more balanced among residential, commercial and industrial types of loads; T4 is dominated by industrial load and is less often encountered in the studied planning region; T5 is a very uncommon case as it has a rarely seen load composition (high residential and industrial). Furthermore, 20 transformers were randomly selected to form up a test set TX (25% of dataset) and their average results are included in Table I.

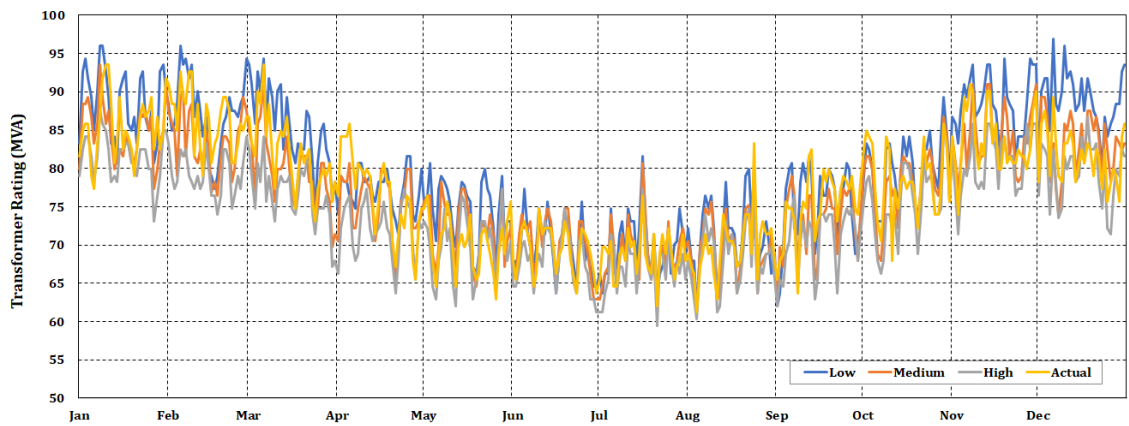

Fig. 8. T1's 2018 Actual Dynamic Rating vs. Estimated Dynamic Rating (with three produced temperature scenarios in 2018)

TABLE I: VERIFICATION RESULTS WITH ACTUAL TEMPERATURE PROFILE

| ID | Area Load Composition $(R, C)$ | Average Membership Probability | Winter Error (%) | | | Summer Error (%) | | |
|---|---|---|---|---|---|---|---|---|
| | | | $ME$ | $AE$ | $VE$ | $ME$ | $AE$ | $VE$ |
| T1 | (3%,88%) | 91% | 5.6 | 2.0 | 4.9 | 5.7 | 2.3 | 4.8 |
| T2 | (57%,43%) | 95% | 5.1 | 3.2 | 6.6 | 4.6 | 3.5 | 5.7 |
| T3 | (24%,34%) | 88% | 4.0 | 2.2 | 3.5 | 5.3 | 1.8 | 6.4 |
| T4 | (0%,12%) | 81% | 10.1 | 5.4 | 11.3 | 9.5 | 6.2 | 7.9 |
| T5 | (36%,4%) | 42% | 8.1 | 5.9 | 6.2 | 9.4 | 4.0 | 8.8 |
| TX | (33%,49%) | 92% | 4.8 | 2.5 | 5.0 | 5.4 | 3.1 | 5.2 |

A number of observations can be made from Figure 7 and Table I:

- In general, daily error $ME$ is much higher than statistical error $AE$. This is expected as there is more fluctuation on a daily basis. This result is positive because $AE$ is more important than $ME$ for long-term planning application. $VE$ as the minimum rating is usually recorded on the days with high temperature or long-lasting high power consumption. It can be either higher or lower than $ME$.

- Transformers T1 to T3 show very good estimation accuracy. They have low industrial load components and high average membership probabilities. They are all common types of transformers in the planning region.

- Transformer T4 shows a relatively lower accuracy. After further investigation, it is found that the major load on T4 is a steel factory that not only operates during the night (by night shift operation) but also often operates over the weekend time. This explains the reason of T4's lower accuracy: its abnormal type of operation makes the load shape profile become very unique. The load shape profile cannot be constructed as accurately as common transformers by collaboratively leveraging the data of other existing transformers in the planning region.

- Transformer T5 shows a relatively lower accuracy. It has a quite low Average Membership Probability. This means that this transformer has a rarely seen load composition (high residential and industrial). In other words, during clustering, this transformer data point is far away from any clusters of existing transformers. This explains the reason of T5's lower accuracy: no existing transformer profile can dominantly approximate T5's load shape profile. Therefore, larger errors could occur when constructing T5's load shape profile and in the subsequent process.

- The average performance of the test set TX is closer to T1 to T3 because they are common types of transformers in the planning region.

The above observations discovered an important caveat for the application of the proposed approach: directly applying the proposed approach to a particular transformer that mainly supplies an irregular type of industrial load or has a unique load composition with respect to other transformers in the same planning region may lead to less accurate estimation. This is because the concept of the proposed GMM based load shape profiling method is similar to the collaborative filtering algorithms used widely today in machine learning for the development of recommendation systems [25]. It works well when similar members present and works less accurately when no similar member presents. As a data-driven approach, this is a limitation to the discussed application. However, the results are still much more accurate than the current nameplate based rating methods which completely disregard the use of any long-term historical information in the planning region. Some practical suggestions are further discussed in Section VI to account for this limitation. When applying to common transformers, the accuracy is quite satisfactory as indicated by the results of T1 to T3.

### B. Verification with Produced Temperature Profiles

The second step of verification focuses on the established long-term temperature profiles as discussed in Section II. Please note that we should respect a basic fact that practically it is impossible to forecast long-term temperatures accurately on a daily basis [10]. The true purpose of establishing long-term temperature profiles is to determine a statistically representative band for the estimation, with a certain adjustment to reflect the long-term trend such as global warming. From the band defined by the best scenario (low-temperature), the worst scenario (high-temperature) and the medium scenario (medium-temperature), utility planning engineers can refer to the results and understand the flexibility and constraints they have when sizing the transformer. The 2018 actual and estimated dynamic rating values of transformer T1 under three temperature scenarios produced as per Section II-A are plotted in Figure 8. The results of common type transformers T1 to T3 as well as the average results of the test set TX are summarized into Table II.

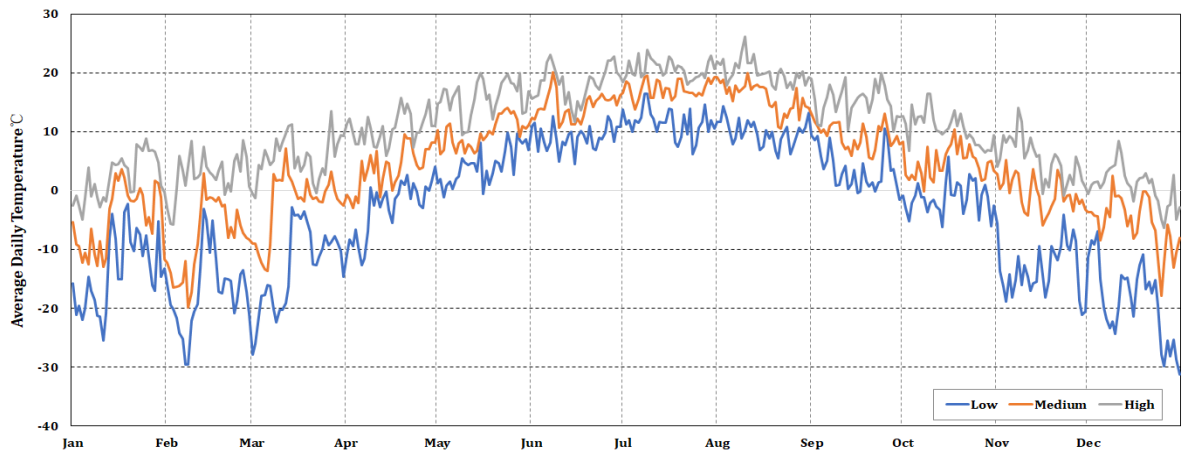

Fig. 9. Established long-term annual temperature profiles based on 2014-2018 data

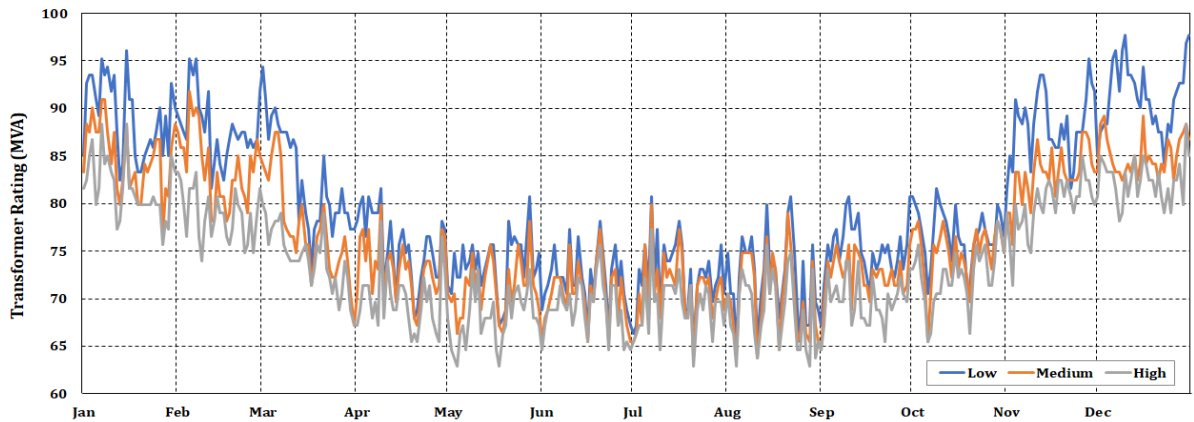

Fig. 10. Forecasted annual dynamic rating profiles in 2023 for a hypothetical transformer

TABLE II: VERIFICATION RESULTS WITH PRODUCED TEMPERATURE PROFILES

| ID | Hypothetical Temperature Senario | Winter Error (%) | | | Summer Error (%) | | |
|----|----|----|----|----|----|----|----|
| | | ME | AE | VE | ME | AE | VE |
| T1 | High | 8.0 | 7.1 | 9.2 | 9.4 | 6.6 | 10.7 |
| | Medium | 6.1 | 4.5 | 5.3 | 6.3 | 3.0 | 5.2 |
| | Low | 8.5 | 5.6 | 7.5 | 8.2 | 5.3 | 6.9 |
| T2 | High | 8.6 | 5.9 | 11.3 | 9.4 | 7.5 | 10.8 |
| | Medium | 6.4 | 4.6 | 7.7 | 6.8 | 6.1 | 7.8 |
| | Low | 9.5 | 6.4 | 9.2 | 8.9 | 4.8 | 8.0 |
| T3 | High | 7.4 | 3.7 | 5.7 | 8.0 | 4.6 | 6.2 |
| | Medium | 4.7 | 3.1 | 4.3 | 5.6 | 3.3 | 7.0 |
| | Low | 5.9 | 3.5 | 4.7 | 7.1 | 5.1 | 4.8 |
| TX | High | 8.2 | 5.3 | 8.6 | 9.0 | 6.0 | 9.5 |
| | Medium | 5.4 | 3.9 | 5.8 | 6.5 | 4.3 | 6.4 |
| | Low | 7.7 | 4.9 | 6.9 | 8.1 | 5.1 | 6.6 |

As can be seen from Figure 8 and Table II, in general, the accuracy is quite satisfactory, especially with the medium temperature scenarios. However, this does not mean the high and low temperature scenarios are not useful and should not be considered in certain cases: first of all, their results are still much more accurate than using nameplate ratings and are within an acceptable range; second, utility engineers may intentionally choose the high-temperature scenario to account for future uncertainties and risks such as global warming or underestimated load growth. Using high-temperature scenario will guarantee a reasonably conservative result and the conservativeness is much lower than relying on nameplate rating. On the other hand, utility engineers may also choose the low-temperature scenario when other operational flexibilities such as feeder ties, load shedding and demand response are available. These measures can allow temporary load transfer to other transformers or loading alleviation during emergency conditions. It can also be considered when load growth in the planning region has been traditionally overestimated.

### C. Future Annual Temperature Profiles

By using the method discussed in Section II and the 5-year historical weather data from 2014 to 2018, three long-term annual temperature profiles for the planning region are established as shown in Fig.9. The historical weather data was obtained from [26].

### D. Future Annual Dynamic Rating Profiles

By using the method discussed in Section III and IV and the 5-year historical weather and loading data from 2014 to 2018,

three annual dynamic rating profiles in 2023 with an assumed future load composition (60%, 30%) for a 50MVA ONAF power transformer typically used by the utility is given in Fig.10. As can be seen, the summer months (May to Sep.) have lower rating than winter months (Oct. to Apr.) and this is because summer has higher ambient temperatures. Also, the high temperature scenario yields low dynamic rating and vice versa.

### E. Sensitivity Analysis by Load Composition

Sensitivity analysis was also applied to analyze transformer ratings for different area load compositions. 4 area load types- residential heavy, commercial heavy, industrial heavy and balanced were considered. Load compositions for each area load type in 2023 were assumed and listed in Table I. A typically used 50MVA ONAF type power transformer is considered. It is discovered that residential heavy and commercial heavy load types have relatively higher ratings while industrial heavy and balanced load types have lower ratings. This is because the industrial loads in the planning region do not fluctuate dramatically as residential and commercial loads in a day and often operate constantly at a high level. This kind of load behavior affects the cooling of transformer temperature.

TABLE III: SENSITIVITY ANALYSIS BY LOAD COMPOSITION

| Area Load Type | Area Load Composition $(R, C)$ | Temperature Scenario | Summer Average Rating (MVA) | Winter Average Rating (MVA) |
|---|---|---|---|---|
| Residential Heavy | (80%,10%) | High | 68 | 78 |
| Commercial Heavy | (10%,80%) | High | 67 | 72 |
| Industrial Heavy | (10%,10%) | High | 64 | 68 |
| Balanced | (33.3%, 33.3%) | High | 65 | 71 |
| Residential Heavy | (80%, 10%) | Medium | 71 | 81 |
| Commercial Heavy | (10%, 80%) | Medium | 70 | 76 |
| Industrial Heavy | (10%, 10%) | Medium | 66 | 71 |
| Balanced | (33.3%, 33.3%) | Medium | 67 | 74 |
| Residential Heavy | (80%, 10%) | Low | 72 | 85 |
| Commercial Heavy | (10%, 80%) | Low | 72 | 79 |
| Industrial Heavy | (10%, 10%) | Low | 67 | 75 |
| Balanced | (33.3%, 33.3%) | Low | 68 | 78 |

### F. Implications for Utility Long-term Planning

The above results showed the great value of the proposed method for utility long-term planning. To determine the proper size of a new power transformer, planning engineers first need to forecast the load compositions or the change of load compositions in the area to be supplied by the transformer over the next few years. This can be typically done by analyzing area development plan and area land characteristics [1]. If facing uncertainty, it is also reasonable to assume different load composition scenarios. Then the annual dynamic ratings of the transformer can be estimated using the proposed method for the next few years. In parallel with the above process, planning engineers will forecast the loading growth for the next few years (often split to summer/winter or quarterly forecasting seasons). The forecasted loading can be compared with the estimated power transformer dynamic rating to determine: the proper size of a new transformer, the need of upgrading an existing transformer to a larger size or the timing of such installation or upgrade. In this analysis process, according to the utility company's risk tolerance level and planning practice, planning engineers can also assume proper global temperature adjustment, select a certain temperature scenario out of the three or produce results under all three scenarios for further cost-risk comparison and sensitivity evaluation. Although the utility long-term planning process can never be 100% accurate, the proposed method can provide in-depth information required to support more scientific and realistic planning decision making. It should also be noted that the proposed method is based on the theoretical transformer thermal model defined in IEEE std. C57.91-2011 which only considers load profile, ambient temperature and typical manufacturing parameters as model inputs. This is a practical standard that has been used in numerous studies and are therefore adopted in here for long-term planning purpose. In reality, if the utility company is concerned about other factors such as manufacturing differences, additional safety margin can be applied to the obtained rating results to account for such uncertainties.

## VI. CONCLUSIONS AND DISCUSSIONS

This paper addresses an important problem in utility companies that has not been researched before – how to produce annual continuous dynamic rating of power transformers for long-term planning purpose. To respond to this need, this paper proposes a novel and comprehensive data analytics method to process the past 5-year temperature, loading and load composition data of existing power transformers in a planning region. The outcomes of the proposed method include:

- Three long-term annual temperature profiles for the planning region can be established;
- For any day in a year, a future power transformer's load shape profile can be constructed by using Gaussian Mixture Model and Silhouette analysis;
- A power transformer thermal aging model can be established with respect to IEEE std. C57.91-2011. Future load shape and temperature profiles under different scenarios can be incorporated into such model and the corresponding transformer rating can be determined;
- Three annual continuous dynamic rating profiles of the future transformer can be produced under three long-term temperature scenarios.

The novelties and significances of this paper can be summarized into two main points as follows:

1) This paper introduces the concept of dynamic rating into the long-term planning process. Previously, there has been a lack of research attention on this subject. The current rating method used for long-term planning is too conservative and often leads to over investment. This paper aims to draw research and application attention to this problem and has a significant economic implication to utility companies (power transformers are very costly components).

2) The paper presents a novel big data approach with sophisticated data analytics techniques to solve the problem. This whole analytics process (establishing long-term temperature profile, finding similar historical days and using probabilistic clustering and Silhouette analysis) is completely novel and unique.

This paper also presents the details of an application example for a major utility company in Canada. It analyzes the validity of the proposed method and explains how such results can help utility planning engineers with long-term system planning. Overall, it demonstrates great practical value and feasibility of the proposed method in real world.

The results show that the estimation accuracy for majority of transformers is satisfactory. However, the accuracy can be affected when dealing with special transformers with unique load composition or load shape that is rarely encountered before in the planning region. Since the method is based on collaboratively leveraging existing power transformer data, it is suggested to apply the proposed method to planning regions with sufficient number of existing power transformers. For example, it is probably not a great idea to apply the method to a small town system with only 4 to 5 power transformers. Finally, when dealing with special transformers with unique load composition or load shape in real application, there are a few practical suggestions which can potentially reduce the impact and improve the estimating accuracy:

1) Apply a higher safety factor and more conservative temperature assumptions when choosing the size of the transformer;

2) Increase operational flexibility to account for unexpected transformer undersizing in the future. Oftentimes this can be more economical than increasing the size of a power transformer. Operational measures such as adding feeder ties to the feeders supplied by the transformer, load shedding and demand response can be adopted;

3) Select the transformer model that has better heat dissipation and oil flow convection capability such as oil-directed air-forced transformers. This can increase transformer's overloading capability in unexpected loading conditions;

4) Large utility companies can consider establishing a database to store transformer profiles with special industrial loads so that they can be referenced in future planning work for a different planning region to correct the load shape estimation.

Future research on this subject could expand to medium or low voltage service transformers that are widely used in distribution systems. With the use of long-term smart metering data, weather data and data analytics, proper transformer fleet asset management strategies can be studied accordingly. Other statistical and modelling methods can also be explored.

**Ming Dong** (S'08, M'13, SM'18) received his Ph.D degree from Department of Electrical and Computer Engineering, University of Alberta. Since graduation, he has been working in various roles at major electric utility companies in West Canada as a Senior Engineer for 7 years. He was the recipient of the Certificate of Data Science and Big Data Analytics from Massachusetts Institute of Technology. He is an associate editor of CSEE Journal of Power and Energy Systems and an editor of IEEE Open Access Journal of Power and Energy. His research interests include applications of artificial intelligence and big data technologies to power system planning, operation and asset management, power quality and smart energy management.